\documentclass[aps,pre,preprint,showpacs]{revtex4-1}%
\usepackage{amsfonts}
\usepackage{amsmath}
\usepackage{amssymb}
\usepackage{epsf, epsfig, graphics}
\usepackage{graphicx}%
\usepackage{color}
\begin{document}
\title{Hydrodynamics of stratified epithelium: steady state and linearized dynamics}
\author{Wei-Ting Yeh$^{1,2}$ \footnote{wtyeh1989@gmail.com} 
 Hsuan-Yi Chen$^{1,2,3}$ \footnote{hschen@phy.ncu.edu.tw}}
 \affiliation{$^{1}$Department of Physics, National Central University, Jhongli, 32001, Taiwan\\
 $^{2}$Institute of Physics, Academia Sinica, Taipei, 11529, Taiwan\\
 $^{3}$Physics Division, National Center for Theoretical Sciences, Hsinchu, 30013, Taiwan}	 				
 \begin{abstract}		
A theoretical model for stratified epithelium is presented.   The viscoelastic properties of the tissue is assumed to be dependent on the spatial distribution of proliferative and differentiated cells.   
Based on this assumption, a hydrodynamic description for tissue dynamics at long-wavelength, long-time limit is developed, and the analysis reveals important insight for the dynamics of an epithelium 
close to its steady state.    
When the proliferative cells occupy a thin region close to the basal membrane, the relaxation rate towards the steady state is enhanced by cell division and cell apoptosis.  
On the other hand, when the region where proliferative cells reside becomes sufficiently thick, a flow induced by cell apoptosis close to the apical surface could enhance small perturbations.  
This destabilizing mechanism is general for continuous self-renewal multi-layered tissues, it could be related to the origin of certain tissue morphology and developing pattern.
\end{abstract}	
 \date{\today}
 \maketitle 
    	     	   
    \section{Introduction}		
    \label{s:introduction}
		Biological tissues are viscoelastic.    Concepts developed for describing passive viscoelastic materials have been applied to biological tissues~\cite{Gonzalez-Rodriguez_12, Lecuit_07}.  However, active processes such as forces generated by the cells, cell movement, and cell proliferation make biological tissues different from passive viscoelastic materials.  Thus biological tissues fall into the category of active soft matters~ \cite{Marchetti_modphys_13}.  On the other hand,  junctions between neighboring cells prevent free diffusion of cells in a tissue.  It makes biological tissues different from other active biological systems such as bacterial colonies and cell cytoplasm.  As a result, biological tissues become interesting systems to study complex collective motion and morphogenesis.  For example, it is known that the homeostasis state of a tissue often has undulating surface to increase the contact area with its environment. 
Understanding the formation and maintenance of such spatial structure 
is thus important for fundamental research of active matters and for applied research such as fabricating functional surfaces and flexible electronics~\cite{Yang_10}.
		
		Experiments have demonstrated that at short time scales a tissue is solid-like due to the adhesion junctions between cells~\cite{Perez-Moreno_03}.  Since the residual stresses can relax due to turnover of junction proteins~\cite{Thoumine_06, Lecuit_07} and cell rearrangement induced by cell division and cell apoptosis~\cite{Ranft_10}, a tissue is liquid-like at long time scales~\cite{Bonnet_12, Foty_94, Forgacs_98, Thoumine_97, Lecuit_07, Gonzalez-Rodriguez_12}. This suggests a characteristic time $\tau$ such that at time scales smaller than $\tau$ a tissue behaves like a solid with shear modulus $E$. On the other hand, at time scales larger than $\tau$, a tissue behaves like a fluid with viscosity $\eta$. Typically $E \sim 10^2-10^4$~Pa~\cite{Forgacs_98}, and $\tau \sim 10 - 10^2 $~s~\cite{Forgacs_98, Bonnet_12}, therefore $\eta = E \tau $ is on the order of $10^3-10^6$~Pa~s~\cite{Basan_09}. 
						
		In this study we focus on stratified epithelium, a multi-layered continuous self-renewal tissue~\cite{Rizvi_05}. Epithelium often forms the outermost layer of skin and mucous membrane, and acts as a barrier separating the outside and inside of a multicellular organism~\cite{Campbell_Bio}. On the top of a stratified epithelium is a free apical surface, on the bottom is a basement membrane attached to the underlying connective tissue 
composed primarily of collagen and elastin fibers.    Typically the thickness of stratified epithelium  is $\sim 100 \ \mu {\rm m}$ and the thickness of the underlying connective tissue is $\sim 1 \ {\rm mm}$ ~\cite{Genzer_06}.
The proliferative cells are often localized near the basal lamina, suggesting the existence of a special microenvironment called stem cell niche~\cite{Watt_00, Moore_06}. Above the proliferative cells are terminally differentiated cells.  They are functional cells that do not divide, instead they undergo programed cell death (apoptosis).  Intuitively, different cells should have different mechanical properties.  As supported by recent experiments, stem cells and tumor cells often have smaller stiffness than normal differentiated cells~\cite{Chowdhury_10, Babahosseini_14}.  This suggests that the mechanical properties of a tissue should also depend on local tissue composition~\cite{Ranft_10}.  
The effects of such inhomogeneity on tissue dynamics is the focus of this study. 
			
		Undulating/fingering structures that are important for biological functions are often seen in the apical surface or basal membrane of a stratified epithelium.  
For example, skin wrinkles or folds are formed when a skin is deformed by external/residual stresses with wavelength $ \sim 1\ {\rm mm}$~\cite{Li_soft_12, Cerda_03, Flynn_09, Genzer_06}. 
At length scale $\sim 100 \ \mu{\rm m}$, rete pegs interdigitate with dermal papillae are observed in the interface between skin epithelium and stroma~\cite{Hale_52, Babler_91, Lavker_83}.  Similar structures have also been observed in the epithelium of mucous membrane~\cite{Li_mech_11, Li_soft_12, Li_appl_11, Klein-Szanto_75}. 
Studies have shown that elastic forces could be important during the formation of the aformentioned structures. 
By modeling a stratified epithelium as an elastic solid in contact with the underlying elastic connective tissue, the formation of skin wrinkles was associated with the buckling instability originated from the competition between bending energy of epithelium and stretching energy of connective tissue~\cite{Cerda_03, Genzer_06}.  The rete peg structure was suggested to arise from the incompatible growth of epithelium and connective tissue~\cite{Ciarletta_12}. Similar mechanism can also explain the formation of fingerprint~\cite{Kuchen_05}, crypt~\cite{Hannezo_11}, and surface wrinkling of tumor spheroids~\cite{Ciarletta_13}.  On the other hand, hydrodynamic instabilities are also likely to produce undulating morphologies in stratified epithelium. For example the formation of rete peg structure could arise from proliferation-induced stresses~\cite{Basan_11}, and an essential condition of this result is that the net cell proliferative rate deceases with the distance from basal lamina. Some models considered more details such as cell lineage dynamics~\cite{Ovadia_13}, viscoelasticity of the  underlying stroma~\cite{Risler_13}, and the effect of growth factors inside a tissue~\cite{Ovadia_13, Risler_13}. Overall, several interesting structures and patterns can be understood by a continuum description of tissues without describing the movement of individual cells and biomolecular events in details. The mechanical stresses induced by cell turnover inside a tissue~\cite{Ranft_10, Basan_09} is sufficient to do this job.
					
		The aim of this work is to study theoretically how inhomogeneity of mechanical properties in a tissue affects the tissue dynamics.  
For illustrative purpose we consider a stratified epithelium on a rigid stroma. 
We begin with a simple model in which the mechanical properties of the tissue is assumed to be homogeneous.  With proliferative cells occupying the lower part, and terminal differentiated cells occupying the upper part of the tissue, this model has a homeostasis state with a flat apical surface.   The linearized dynamics close to the homeostasis state in this model already reveals interesting effects from flows and cell proliferation.   For example, cell division/apoptosis speed up tissue relaxation towards the homeostasis state; but this effect is relatively weak when the wavelength of the perturbation is of the order of the thickness of the tissue. 

Due to frequent cell division events, tissue viscosity in the region rich in proliferative cells should be different from the rest of the tissue.  This feature is included in our second model.  
Such a modification indeed leads to a different dispersion relation for tissue dynamics close to the homeostasis state.  
Especially, when the proliferative region becomes sufficiently thick, cell apoptosis close to the apical region could induce a flow that strongly hinders the relaxation towards the steady state.  This effect is 
especially significant for perturbations with wavelength comparable to tissue thickness ($\sim 100 \ \mu$m for typical stratified epithlium).
Since our model is quite general, we expect this destabilizing mechanism to exist in continuous self-renewal tissue, but it is not easy to be observed in normal healthy tissue because the proliferative cells only occupy a thin region.   However, during development or regeneration there is often a transient increase of proliferative cell inside tissues \cite{Lander_09}. As a result, the new effect described by our second model could happen, and it is even possible that undulating apical surface could be generated by this mechanism.

		This article is organized as follows.  Both models are presented in Sec.~\ref{s:method}, where we also discuss the homeostasis states and linearized dynamics close to the steady states.	
In Sec.~\ref{s:discussion} we discuss the flow in an epithelium that is induced by cell division/apoptosis.   Using the result derived in the Appendix, we also discuss the relation between
the dimensionless parameter $\tilde{r} = r_D/r_S$ and $\eta _{rel} = \eta _D/\eta_S$, where $r_D$ and $r_S$ are cell apoptosis rate and cell division rate, and $\eta _D$, $\eta_S$ are viscosity in region rich in differentiated cells and proliferative cells, respectively.  In Sec.~\ref{s:conclusion}, based on these results, we make a conclusion and propose possible experimental tests to validate our theory.   The Appendix presents the relaxation rate in our second model, then a derivation of the tissue viscosity from a hydrodynamic model is presented.
		
\section{MODELS} \label{s:method}
\subsection{First model: stratified epithelium with constant viscosity} \label{s:method-model1}				
Consider a basel lamina sitting on the $xy$-plane.  An epithelium grows from this surface into $z>0$ region.  On top of the epithelium is a free apical surface.  We are interested in the steady state of this system and the linearized dynamics close to this steady state.
Since epithelial growth typically takes several days to complete \cite{Babler_91, Hale_52}, in this article we focus on the long time behavior of our model.  Therefore processes with relaxation time not longer than the time scale $\tau$ introduced in Sec.~\ref{s:introduction} have all decayed away.  The elasticity of the tissue can be neglected, and the  tissue behaves as a viscous fluid \cite{Forgacs_98, Foty_94, Bonnet_12, Thoumine_97}.  The viscosity of the tissue depends on cell-turnover events \cite{Ranft_10} and the rearrangement of junction proteins \cite{Lecuit_07, Thoumine_06}.  Experimentally, the viscosity of a tissue can be estimated from the elastic constant $E$ and relaxation time $\tau$ through the relation $\eta = E \tau$.
		
		Since the dynamics is slow, the inertia of the tissue is also negligible.  Imposing incompresibility for simplicity, the equation of continuity, force balance, and consitutitive equation are 
\begin{equation}\label{gov_cont}
	\partial_l v_l = k_p,
\end{equation}
\begin{equation}\label{gov_fbalance}
	\partial_i \sigma_{ik} = 0,
\end{equation}
\begin{equation}\label{gov_conrel}
	\sigma_{ik} = -p \delta_{ik} + 2 \eta v_{ik}.
\end{equation}
Here $v_i$ is the $i$-th component of the flow velocity field, $v_{ik} \equiv (\partial_i v_k + \partial_k v_i)/2$ is the strain rate tensor in the tissue, $p$ is the tissue pressure, $\eta$ is the viscosity of the tissue, and $k_p$ is the (net) cell proliferation rate in the tissue.  In general $k_p$ is a decreasing function of $z$ because both nutrient and proliferative cells are relatively abundant near the basal lamina.  In this model $\eta$ is chosen to be a constant for simplicity.  However, in general $\eta$ should depend on local tissue composition through forces generated during cell division and cell apoptosis~\cite{Ranft_10}, this will be considered in Sec.~\ref{s:method-model2} when we introduce our second model.
			
			We focus on the situation when the tissue is soft compared to the stroma, therefore the stroma is modeled as a rigid substrate for simplicity, and the flow field vanishes at $z=0$.  More general model that includes the viscoelastic properties of the stroma will be deferred to future studies. The apical surface of the tissue is assumed to be free, and the free boundary condition that includes contribution from surface tension of the tissue and external pressure is applied.
			
			The steady state velocity field of Eqs.~(\ref{gov_cont})-(\ref{gov_conrel}) with these boundary conditions is $v^*_x = v^*_y = 0$ and $v^*_z (z)= \int_0^z k_p(z')dz'$.  The superscript $*$ is used to denote the steady state values.  Since $v_z^*$ should vanish at the apical surface, we have the following steady state condition
\begin{equation}\label{ssc}
	0 = \int_0^{H^*} k_p(z)dz,
\end{equation}
where $H^*$ is the thickness of the tissue in the steady state. Since $k_p$ is a decreasing function of $z$, for the solution of Eq.~(\ref{ssc}) to exist, the tissue needs to have $k_p(0) > 0$ and $k_p(H^*) < 0$. Intuitively, this means that cell division events in the lower part of the tissue should be balanced by cell apoptosis in the upper part of the tissue. We emphasize that our model is independent of the details of $k_p(z)$, the only constraint for $k_p(z)$ is to have steady-state tissue height $H^* >0$ which satisfies Eq.~(\ref{ssc}). 
			
			A homeostasis state of a tissue is a steady state that is stable against small perturbations.  To check if the steady state in this model corresponds to a homeostasis state, we consider a perturbation that takes the tissue height from $H^*$ to $H(x,t) = H^\ast + \delta H(x,t)$.  Similarly, $v_k = v_k^\ast + \delta v_k$, $p = p^\ast + \delta p$, etc.  The linearized equations for the perturbed system are
\begin{equation}\label{pe_cont}
	\partial_l \delta v_l = 0,
\end{equation}
and
\begin{equation}\label{pe_fbalance}
	-\partial_k \delta p + \eta \partial_l \partial_l \delta v_k = 0.
\end{equation}
Note that the perturbed tissue is translationally invariant in the $y$-direction.
			
			For a perturbation with $\delta H, \delta v_i, \delta p \sim e^{iqx+\omega t}$.  The no-slip boundary condition at $z = 0$ gives $\delta v_x|_{z=0} = \delta v_z|_{z=0} = 0$.  The linearized boundary conditions at $z = H^*$ are
\begin{equation}\label{pe_bb1}
		-\delta p |_{z=H^\ast} + 2 \eta \partial_z \delta v_z |_{z=H^\ast} = \sigma \partial_x \partial_x \delta H,
\end{equation}
and 
\begin{equation}\label{pe_bb2}
		\eta\left( \partial_z \delta v_x |_{z=H^\ast} + \partial_x \delta v_z|_{z=H^\ast} \right) = -2\eta k_p(H^*)\partial_x \delta H,
\end{equation}
where $\sigma$ is the surface tension of the apical surface. The above two equations describe the stress balance in the normal and tangential directions of the apical surface. There is also a linearized kinematic boundary condition connecting the evolution of the tissue surface and the flow field
\begin{equation}\label{pe_bb3}
		\partial_t \delta H = \delta v_z |_{z=H^*} + k_p(H^*) \delta H.
\end{equation}
			
\begin{figure}[t]
\begin{center}
	\epsfxsize= 6.5 in \epsfbox{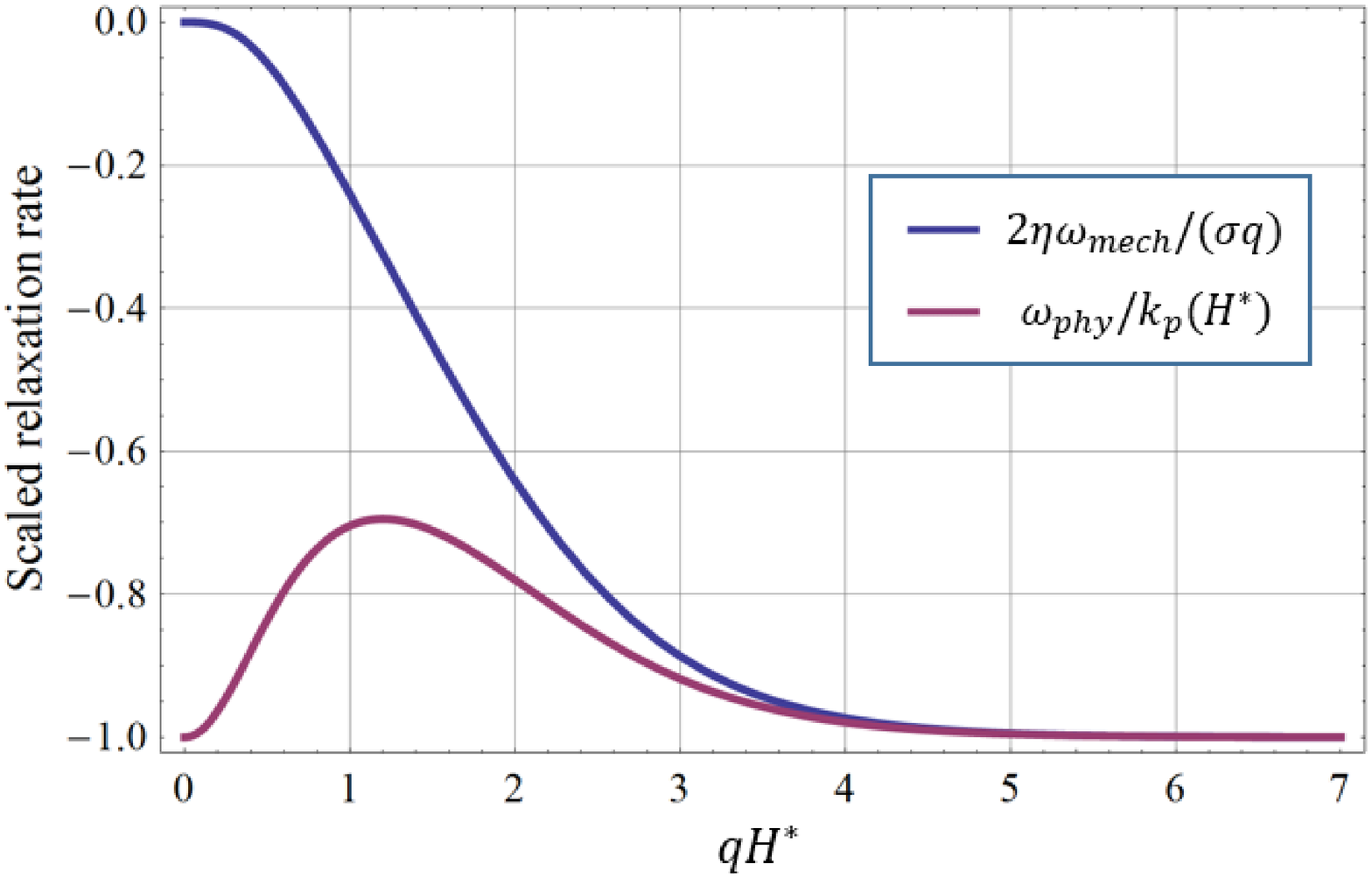}
	\caption{ $2 \eta \omega _{mech}/ \sigma q  $ (blue curve) and  $ \omega _{phy}/ k_p(H^\ast)$ (red curve).
$\omega _{mech}(q)$ decreases monotonically towards $\sigma q / 2 \eta$ as $q H^{\ast} \gg 1$, $| \omega _{phy} |$ has a minimum at $qH^\ast \sim 1$. }  
	\label{f:mech_unsphy_contribution}
\end{center}				
\end{figure}

	The dispersion relation $\omega (q)$ can be calculated from the linearized equations and  boundary conditions of the perturbed system, straightforward algebra gives 
\begin{equation}\label{dispersion}
	\omega (q)= \omega_{mech} (q) + \omega_{phy} (q).
\end{equation}
The first term 
\begin{equation}\label{mech_con}
	\omega_{mech} \equiv \frac{2qH^\ast - \sinh 2qH^\ast}{2(qH^\ast)^2+(1+\cosh 2qH^\ast)} \frac{\sigma q}{2 \eta}
\end{equation} 
is the same as the dispersion relation for the surface of a low Reynolds number simple fluid.   In the limit of large $q H^\ast$ this term approaches $\sigma q / 2 \eta$ \cite{ref:Wu_PRA_1973}.  As Fig.~\ref{f:mech_unsphy_contribution}~(a) shows, $\omega_{mech}$ is negative for all $q$, this means that the surface tension of the apical surface tends to stabilize the steady state.  The second term is
\begin{equation}\label{phy_con}
	\omega_{phy} \equiv  \frac{1+\cosh (2qH^\ast)}{2(q H^\ast)^2+ 1+\cosh(2qH^\ast)} \  k_p(H^\ast).
\end{equation}
It is always negative since $k_p(H^*)<0$, i.e., the apical region of the tissue is rich in terminal differentiated cells which do not divide but undergo apoptosis.  The meaning of this term is intuitive:  cell proliferation and cell apoptosis help to bring a perturbed tissue back to the steady state.	     Note that for given $k_p(H^\ast)$,  $\omega_{phy}$ has a minimum at wavenumber $q = \Xi / H^\ast$, where $\Xi$ is the solution of  $\cosh \Xi = \Xi \sinh \Xi$. Numerically one finds that $\Xi \cong 1.2$ and $\omega_{phy} (q= \Xi/H^\ast) \cong 0.7 k_p(H^\ast)$.

		In this simple model, it is convenient to think that the relaxation of the tissue towards the steady state is due to apical surface tension and cell proliferation.  Contribution from surface tension grows as the wavenumber $q$ of the perturbation increases, approaching $\sigma q/2 \eta$ in the limit of large $qH^\ast$.  Contribution from cell proliferation has a minimum at $qH^\ast \sim \mathcal{O}(1)$.

\subsection{Second model: effect of viscosity heterogeneity} \label{s:method-model2}
			In a typical stratified epithelium, proliferative cells are located close to the basal region~\cite{Watt_00, Moore_06}. Due to their small abundancy, typically the division rate of proliferative cells is large compared to the apoptosis rate of differentiated cells~\cite{Wu_03}. Since stress relaxation in a tissue strongly depends on the division and apoptosis of the cells~\cite{Ranft_10}, the viscosity in regions rich in proliferative cells should be smaller than the viscosity in the other regions of the tissue. Detailed discussion of the relation between cell division/apoptosis and tissue viscosity is presented in Sec.~\ref{s:result-cell composition} and the Appendix.		
		
			To consider position-dependent viscosity we modify our model in Sec.~\ref{s:method-model1} by assuming that all the proliferative cells reside in the region $z < h_S$, and differentiated cells are located at $z > h_S$. 
This is clearly an over simplification, nevertheless it captures the basic tissue composition in a  stratified epithelium.  The net cell proliferation rate inside the tissue is simply  $k_p = r_S$ for $z < h_S$ and $k_p = -r_D$ for $z > h_S$. Here $r_S$ is the cell division rate and $r_D$ is the cell apoptosis rate. From Eq.~(\ref{ssc}), the proportion of proliferative region in the steady state is related to $r_S$ and $r_D$ by
\begin{equation} \label{ssc-model2}
	\frac{h_S}{H^\ast} = \frac{\tilde{r}}{1 + \tilde{r}},
\end{equation}
where $\tilde{r} \equiv r_D/r_S$. We emphasize that the control parameter in typical experiments is  $\tilde{r}$.  For example, $\tilde{r}$ can be increased by increasing the apoptosis rate of differentiated cells \cite{Holcomb_95}. For given $\tilde{r}$ and $h_S$, we can use Eq.~(\ref{ssc-model2}) to obtain the steady state tissue height $H^\ast$. As a reminder, note that in this model $k_p (H^\ast) = -r_D$.
			
     With the aforementioned simplification, in this model we choose $\eta = \eta_S$ for $z < h_S$ and $\eta = \eta_D$ for $z > h_S$. 
The geometry and boundary conditions for the linearized tissue evolution equations are the same as those for our first model, except for additional velocity and stress continuity conditions at $z = h_S$.   The steady state of our second model therefore becomes $v_x^*=v_y^*=0$, and $v_z^* = r_S z$ for $z < h_S$, $v_z^*= r_S h_S - r_D z$ for $h_S <z < H^*$.  The continuity of normal stress at $h_S$ leads to a discontinuity of tissue pressure in our model that should be smoothed out in a real tissue because the boundary between proliferative region and differentiated region is diffuse, not infinitely sharp. 

		On the other hand, in our second model the dynamics close to the steady state are qualitatively different from that of the first model.   Straightforward calculation for a small perturbation of the form $\delta H \sim e^{iqx + \omega t}$ around the steady state of our second model gives the relaxation rate $\omega$. We find that $\omega$ can still be expressed as the sum of two terms $\omega_{mech}$ and $\omega_{phy}$, but these two terms are different from the counterparts of our first model,
\begin{equation} \label{new_mech} 
     \omega _{mech} = K_{mech} (qH^*, \tilde{r}, \eta _{rel}) \frac{\sigma q}{2 \eta_D},
\end{equation}						
\begin{equation} \label{new_phy}
	\omega_{phy} = K_{phy} (qH^\ast,\tilde{r},\eta_{rel}) ~r_D.		
\end{equation}
The explicit expression of  the dimensionless functions $K_{mech}$ and $K_{phy}$ are pretty lengthy, we present them in the Appendix.
These functions are plotted in Fig.~\ref{f:new_mech_phy} as these figures help us to get a better idea of how they behave as the parameters $\eta _{rel}$, $qH^*$ and $\tilde{r}$ change.  Besides their shapes, it is important to point out two issues.  First, $\omega _{mech}$ is the same as the dispersion relation of the upper surface of a phase-separated binary low Reynolds number fluid with lower fluid viscosity $\eta _S$, thickness $h_S$, and upper fluid viscosity $\eta _D$, thickness $H^*-h_S$.  
Second,  in the limit  $\eta_{rel} = 1$,  Eq.~(\ref{new_mech}) and (\ref{new_phy}) become Eq.~(\ref{mech_con}) and (\ref{phy_con}) as we expected.

	Figure~\ref{f:new_mech_phy} shows $K _{mech}$ and $K _{phy}$ for $\eta _{rel} \neq 1$.  It can be seen that $K_{mech}$ is always negative (stabilizing), and its magnitude increases with $\tilde{r}$.  This is because the thickness of the lower (less viscous) region of the tissue increases and the damping is slowed down as $\tilde{r}$ increases.    
 On the other hand, $K_{phy}$ is a non-monotonic function of $\tilde{r}$, as this term is originated from cell division/apoptosis, it cannot be simply explained by the viscous properties of the tissue.  
 We will discuss more about the behavior of $K_{phy}$ in the next section when the flow field in the tissue is examined.

\begin{figure}[t]
	\begin{center}
		\epsfxsize=6.5in \epsfbox{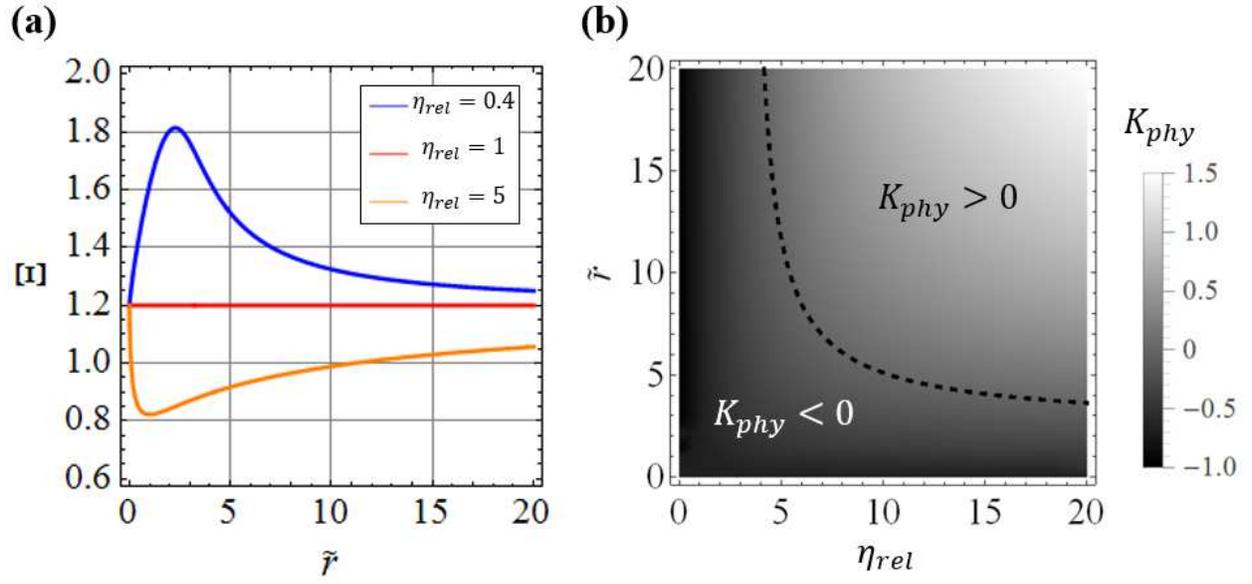}
		\caption {(a) $\Xi$ as a function of $\tilde{r}$ for $\eta _{rel}=0.4$ (blue), 1 (red), and 5 (orange).
		              (b) This plot shows the least negative value of $K_{phy}$ for given $\tilde{r}$, $\eta _{rel}$, i.e., $K_{phy}(qH^*=\Xi, \tilde{r}, \eta _{rel})$.  The dotted curve shows where $K_{phy}(qH^*=\Xi, \tilde{r}, \eta _{rel})$ is zero. }
		\label{f:Xi}
	\end{center}				
\end{figure}

      For fixed $\tilde{r}$ and $\eta _{rel}$,  $K _{phy}$ has a minimum at $qH^{*} = \Xi (\tilde{r}, \eta _{rel})$.  Note that, unlike our first model, here $\Xi$ depends on $\tilde{r}$ and $\eta_{rel}$.  
Figure ~\ref{f:Xi}(a) shows $\Xi$ versus $\tilde{r}$ for different choices of $\eta _{rel}$.   
It is clear that in our second model the minimum of $\omega _{phy}$ also occurs at $qH^* \sim \mathcal{O}(1)$.  
Figure~\ref{f:Xi}(b) shows that when $\tilde{r}$ and $\eta _{rel}$ are large, it is possible for $K_{phy}$ to become positive.  That is, it is possible for cell division/apoptosis to either slow down the relaxation toward the steady state, or even drive the tissue to an instability.  This will be discused in details in Sec.~\ref{s:result-double edged}.
					 
In this section $\eta_{rel}$ and $\tilde{r}$ are treated as independent variables.  We will discuss the relation between $\eta _{rel}$ and $\tilde{r}$ in Sec.~\ref{s:result-cell composition}.
			
\section{RESULTS AND DISCUSSIONS} \label{s:discussion}
\subsection{Flow induced by cell apoptosis on the apical surface} \label{s:result-double edged}    				
					An immediate interesting question of our hydrodynamic theory of epithelium tissue is, besides fluidizing the tissue, how cell turnover inside the tissue affects tissue dynamics.  This is considered in both our models.  In the first model, we have assumed that the net cell proliferation rate in a tissue is the only property that depends on the distance from the basal membrane, all other properties of the tissue are homogeneous.  On the other hand, in the second model, regions with different net cell proliferation rate are assumed to have different viscosity.   Interestingly, these two models predict very different tissue dynamics, revealing nontrivial effects of cell turnover on tissue dynamics.
					
					In our first model (Sec.~\ref{s:method-model1}),  the tissue is homogeneous except for the proliferative property.  The linearized dynamics near the steady state shows that the steady state is always stable under small perturbations.  The relaxation rate towards the homeostasis state contains two terms (Eq.~(\ref{dispersion})). The first term $\omega_{mech}$ (Eq.~(\ref{mech_con})) is driven by the surface tension of the apical surface, it is the same as the dispersion relation for the liquid-air surface of a low Reynolds number simple liquid.  It always stabilizes the tissue and this rate increases with the wavenumber $q$ (Fig.~\ref{f:mech_unsphy_contribution}~(a)). 
					
\begin{figure}[t]
	\begin{center}
		\epsfxsize=6.5in \epsfbox{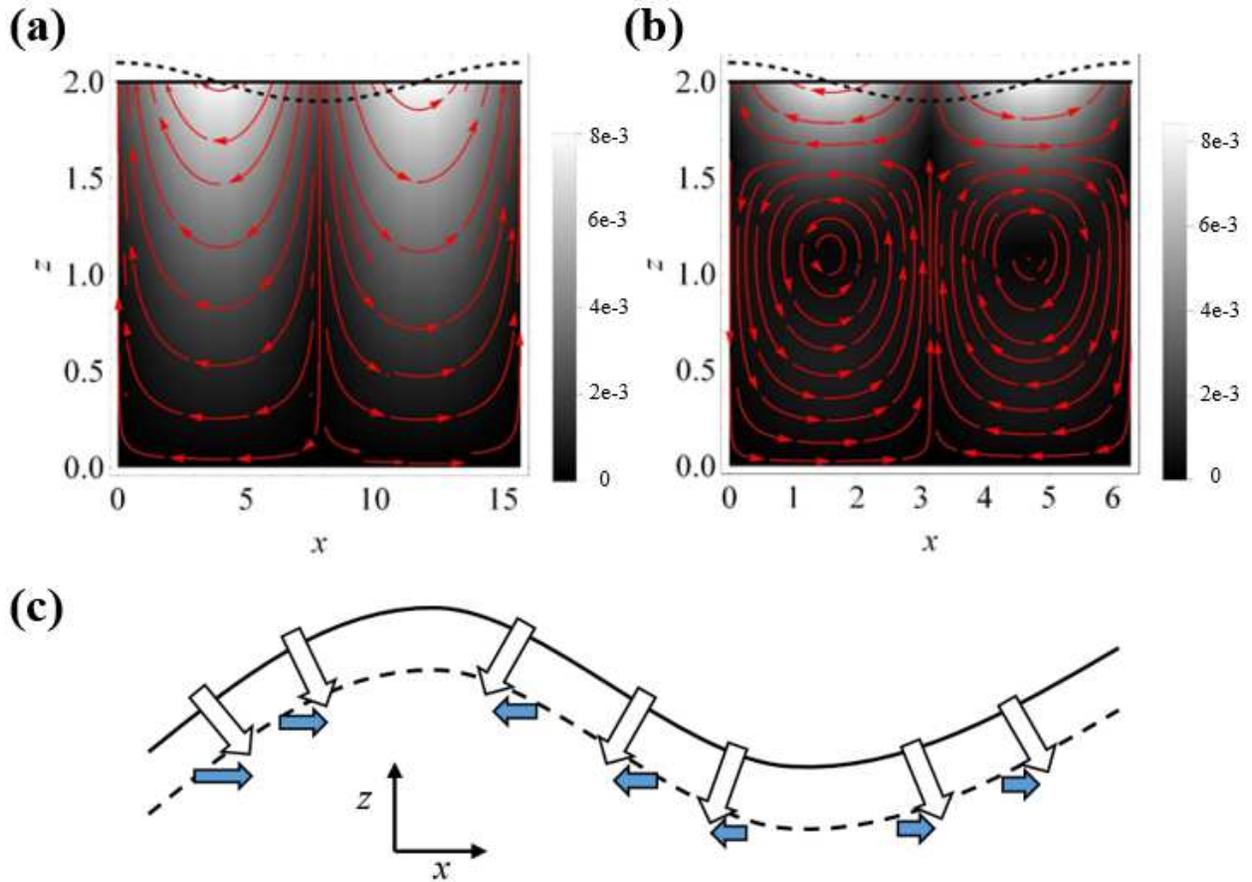}
				\caption{ (a) and (b): Field lines of $(\delta v_x, \delta v_y)$ calculated from our first model for a perturbed tissue with tissue apical surface tension $\sigma = 0$.  It can be seen that the topology of the flow field depends on  the magnitude of $qH^\ast$.  $qH^\ast = 0.8 $ in (a), $qH^\ast = 2 $ in (b). The red arrows are the stream lines, the dashed black lines are the apical surface of the perturbed tissue, and the background gray-level indicates the magnitude of $\delta v$. To generate the flow field, we have chosen $k_p(H^\ast) = -1$, $H^\ast = 2$, and $\delta H = 0.01\cos(qx)$. 
(c) Cell death at apical surface induces a flow field that is indicated by the empty arrows.  Close to the apical surface this flow field has a horizontal component indicated by the blue arrows. The solid curve is the apical surface at the present moment, and the dashed curve indicates the apical surface at a later time. }
				\label{f:model1_flow}
						\end{center}				
					\end{figure}

					The second term $\omega_{phy}$ (Eq.~(\ref{phy_con})) comes from cell turnover in the tissue.  To gain further insight on the effect of this term to the dynamics of the tissue, in Fig.~\ref{f:model1_flow}~(a)(b) the field lines of $(\delta v_x, \delta v_y)$, the deviation of velocity field from the steady state in a tissue is plotted for a tissue with zero apical surface tension $\sigma$ for two different  values of $qH^*$.  Since $\sigma = 0$, the relaxation of the tissue is completely driven by the mechanism represented by $\omega_{phy}$.  It can be clearly seen that there is a flow induced by cell apoptosis at the apical surface.  As shown schematically in Fig.~\ref{f:model1_flow}(c), cell apoptosis close to the apical surface gives an effective surface flow.  This is the mechanism that drives the flow field shown in Fig.~\ref{f:model1_flow}(a) and (b).  Figure~\ref{f:model1_flow}(b) shows that when $qH^*$ is far from $\Xi$ ($\Xi \approx 1.2$), the flow is restricted to region close to the apical surface.  On the other hand, Fig.~\ref{f:model1_flow}(a) shows that when $qH^*$ is closer to $\Xi$, the flow can penetrate deeper into the tissue.  As $q$ further decreases to $qH^\ast \ll 1$ the apical surface becomes nearly flat, and the induced horizontal flow (not shown) is again small.  Therefore the effect of induced flow is significant when $qH^* \approx \Xi$.  This is reflected in Sec.~\ref{s:method-model1} as the minimum of the magnitude of $\omega _{phy}$ at $qH^*$ of order unity.   The fact  that $\omega _{phy}$ has a minimum at $qH^* = \Xi$ also indicates that although intuitively cell turnover helps to stabilize the tissue, flow induced by apoptosis at apical surface slows down the relaxation of the tissue.  Indeed, as can be seen from  Fig.~\ref{f:model1_flow}~(c), the induced flow tends to push cells from regions where the tissue is thin to regions where the tissue is thick.    Anyway, combining these two effects, in this model $\omega _{phy}$  is still negative for all $q$.  That is, our first model predicts that cell turnover helps to stabilize the homeostasis state, although the stabilizing effect is relatively weak when $qH^* \sim 1$.  It is interesting to note that Fig.~\ref{f:model1_flow}(a) and Fig.~\ref{f:model1_flow}(b) also shows that vortices appears in the flow field when $qH^* > \Xi$, and they are absent when $qH^* < \Xi$.
					
\begin{figure}[t]
\begin{center}
\epsfxsize=6.5in \epsfbox{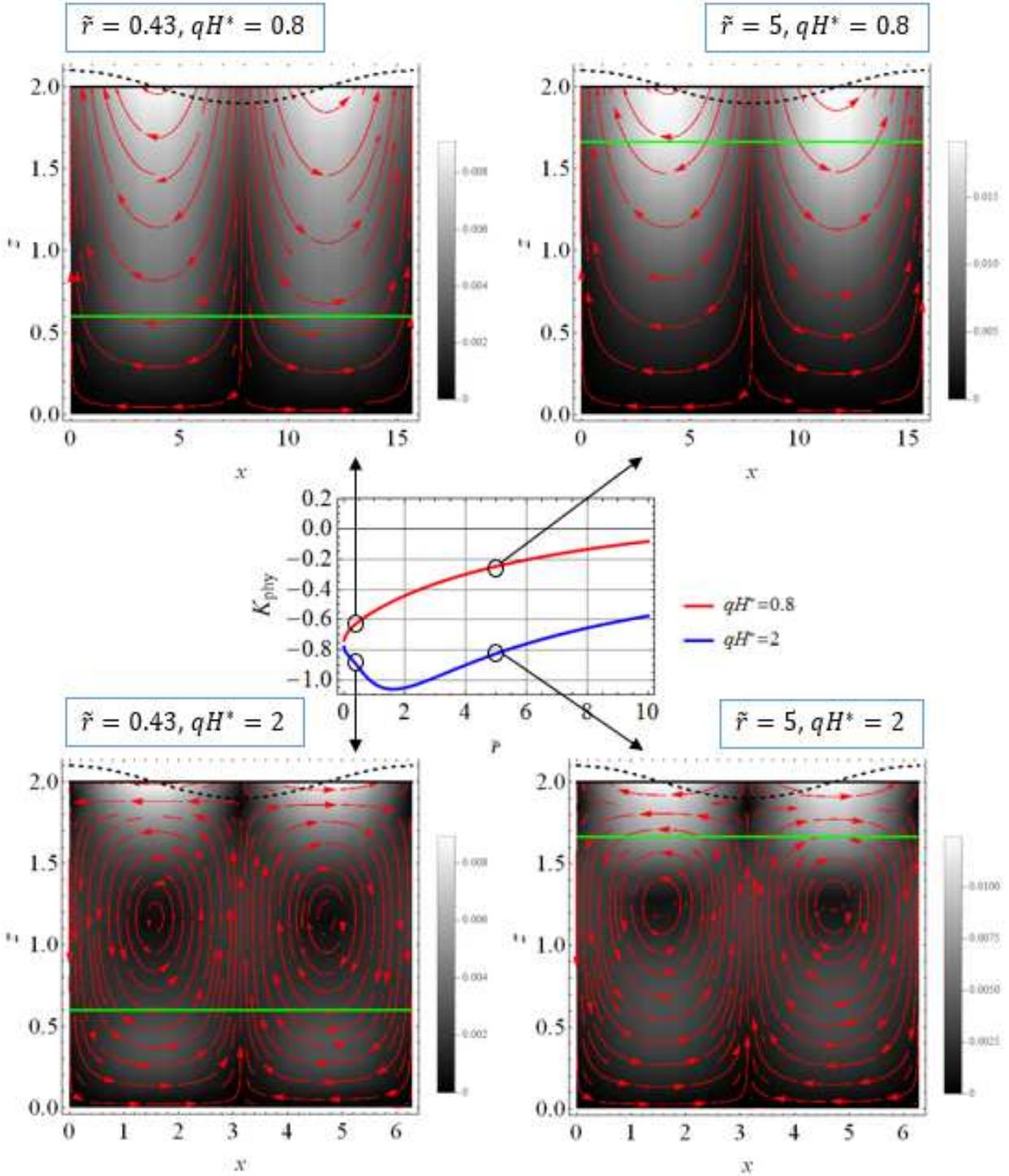}
\caption{Field lines of $(\delta v_x, \delta v_y)$ in the second model for $\sigma =0$, and $K_{phy}$ versus $\tilde{r}$ for $qH^\ast =0.8$ ($< \Xi$ for all $\tilde{r}$) and $qH^* = 2.0$ ($> \Xi$ for all $\tilde{r}$). For $qH^\ast = 0.8$, $K_{phy}$ increases monotonically with $\tilde{r}$. For $qH^\ast = 2$, there are vortices in the induced flow.  $K_{phy}$ decreases with $\tilde{r}$ when $h_c > h_S$, $K_{phy}$ increases with $\tilde{r}$ when $h_c < h_S$.  $h_S$ is indicated by the green lines, and $h_c$ is the $z$-position of the vortex center.  To generate the flow fields, we have chosen $\eta _{rel} = 5$, $k_p(H^*)=-1$, $H^*=2$, and $\delta H = 0.01 \cos (qx)$.}
\label{f:model2_flow}
\end{center}				
\end{figure}

					Fig.~\ref{f:model2_flow} shows the field lines of $(\delta v_x, \delta v_z)$ predicted by our second model for a perturbed tissue with zero apical surface tension.  Again the field lines for $qH^\ast > \Xi$ look very different from the field lines for $qH^\ast < \Xi$.  
As $qH^*$ increases from  $qH^*< \Xi$ to $qH^* > \Xi$, vortices appear in the induced flow, this is similar to our first model (see Fig.~\ref{f:model1_flow}~(b)).  Let $h_c$ be the height of the center of the vortices.   When $qH^* > \Xi$, Fig.~\ref{f:model2_flow} shows that for $h_c > h_S$, $K_{phy}$ decreases as $\tilde{r}$ increases; however, when $h_c < h_S$, $K_{phy}$ increases as $\tilde{r}$  increases. When $h_c= h_S$, $K_{phy}$ has a minimum (not shown in the figure).     
On the other hand, when $qH^*<\Xi$, $K_{phy}$ increases monotonically with $\tilde{r}$.  
This shows that the shape of $K_{phy}$ in Fig.~\ref{f:new_mech_phy}(c) is related to both the flow induced by cell division/apoptosis and the thickness of the proliferative region.  
					
					 A very important result of our second model is that the sign of $K_{phy}$ can be positive, i.e., it is possible that the overall effect of cell division/apoptosis is to hinder the relaxation of the tissue towards the steady state.  This occurs when the following two conditions are both satisfied.  (i) The  viscosity $\eta _S$ in the region $h < h_S$ is sufficiently small compared to $\eta_D$ (the viscosity in the region $z>h_S$).  (ii) The relative thickness of the proliferative region  $h_S/H^*$, is sufficiently large.  In Sec.~\ref{s:result-cell composition} we will discuss the relation between the distribution of proliferative cells and viscosity, there we will show that indeed in general $\eta _{rel} = \eta _D/\eta _S >1$.    However, a tissue that satisfies condition (ii) is likely not a normal tissue, as a healthy mature epithelium tissue usually has a relatively thin layer of proliferative cells, a tissue with large $h_S/H^*$  is more likely to be found during development or regeneration.
					 
\begin{figure}[t]
\begin{center}
	\epsfxsize=6.5in \epsfbox{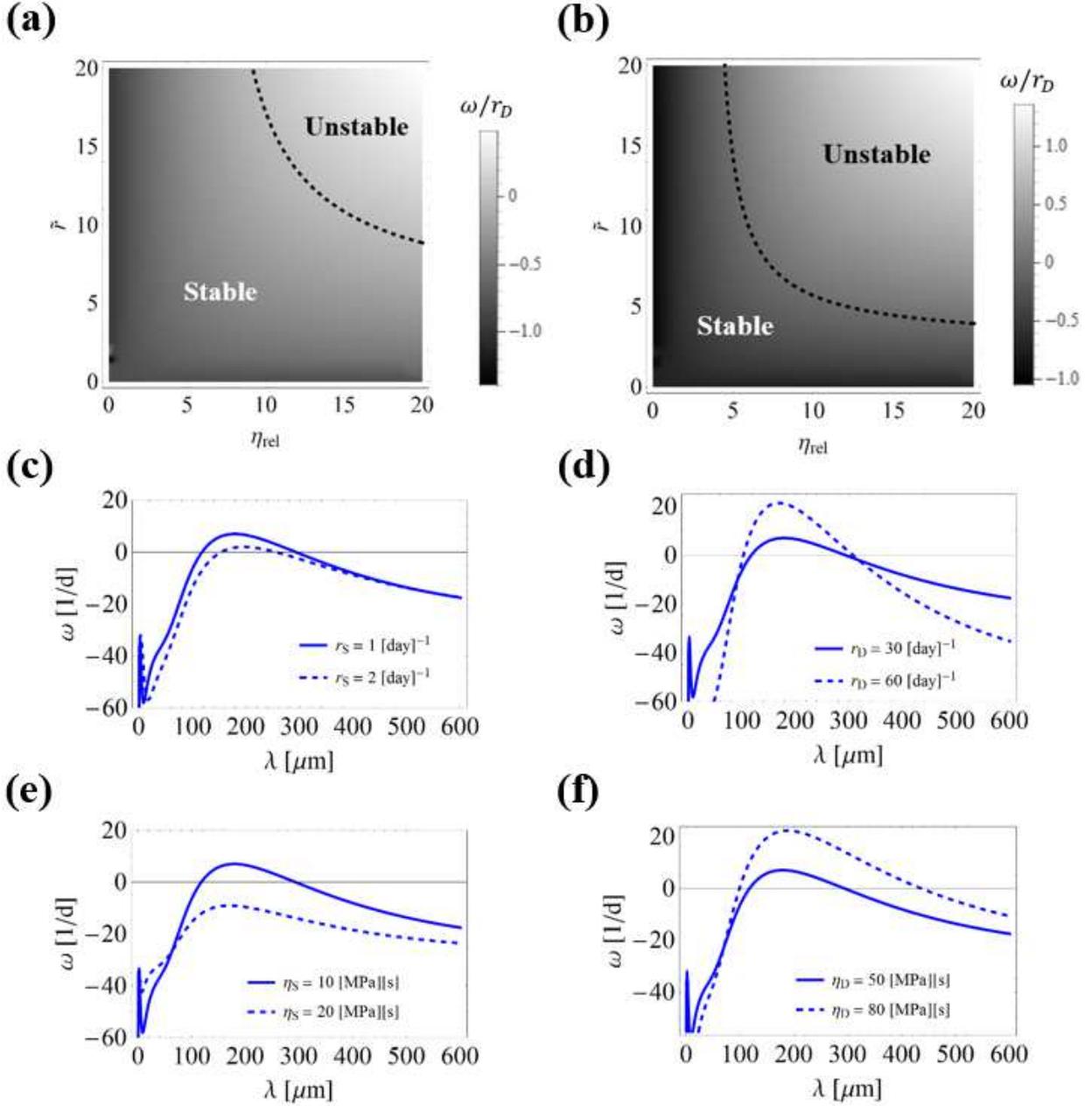}
	\caption{(a)(b) Gray scale indicates th relaxation rate of a perturbed tissue scaled by $r_D$ for $qH^\ast = \Xi$.  (a) $\sigma q / (2 \eta_D r_D) = 0.4$, and (b) $\sigma q / (2 \eta_D r_D) = 0.05$. (c)-(f) relaxation rate (in units of $1/$day) of a tissue as a function of the wavelength of perturbation.   Solid curves:  $r_S = 1~{\rm  d}^{-1}$, $r_D = 30~{\rm d}^{-1}$, $\eta_S = 10~{\rm MPa \ s}$, $\eta_D = 50~{\rm MPa \ s}$, $\sigma = 1~{\rm m \ Nm}^{-1}$, and $h_S =30~\mu {\rm m}$  (the corresponding magnitude of  $H^\ast$ is $31~\mu{\rm m}$).  The dashed lines give the relaxation rate versus $\lambda$ for the same parameters except $r_S = 2 \ {\rm d} ^{-1}$ in (c), $r_D = 43 \ {\rm d}^{-1}$ in (d), $\eta _S = 20 {\rm M Pa \  s}$ in (e), and $\eta _D = 80 {\rm M  Pa \ s}$ in (f). }
	\label{f:model2_diagram}
\end{center}				
\end{figure}

					 Stability diagrams of our second model are shown in Fig.~\ref{f:model2_diagram}~(a) and (b). It is convenient to introduce a dimensionless parameter $\sigma q / (2 \eta_D r_D)$ to describe the relative importance of the apical surface tension and cell turnover in tissue dynamics.  As can be seen by comparing Fig.~\ref{f:model2_diagram}~(a) and (b),  tissues with small $\sigma q / (2 \eta_D r_D)$ are relatively easy to become unstable, this is due to the  relatively weak stabilizing effect from the apical surface tension.
			
			Intuitively cell turnover does not necessarily drive a tissue unstable.  Indeed we know from our previous discussion that sometimes cell turnover helps to stabilize the homeostasis state, while sometimes cell turnover drives a steady state unstable.  Therefore it is interesting to see when cell turnover stabilizes or destabilizes a steady state.  Fig.~\ref{f:model2_diagram}~(c)-(f) shows how growth rate of small perturbations close to the steady state changes with our model parameters. It suggests that increasing $r_S$ and $\eta_S$ helps to stabilize the steady state, as the peak of the growth rate decreases; on the other hand, increasing $r_D$ and $\eta_D$ helps to destabilize the steady state, as the maximum growth rate of small perturbations increases.  This further confirms the intuition suggested by Fig.~\ref{f:model2_diagram}(a) and ~\ref{f:model2_diagram}(b).

Now we can summarize the role of cell turnover in the dynamics of a tissue close to a steady state.  When the tissue height is perturbed from its steady state value, cell turnover drives the tissue toward the steady state.  On the hand, cell apoptosis on the apical surface induces a flow in the tissue that brings more cells to thicker regions and pushes cells away from regions with smaller thickness.  For a normal tissue, proliferative cells occupy a small region close to the basal surface, the overall effect of these mechanisms to tissue dynamics always helps to stabilize the tissue.  On the other hand, for a tissue with a large proportion of proliferative cells, when the viscosity of the proliferative region of the tissue is sufficiently small compare to the viscosity of the non-proliferative region, the induced flow can destabilize the steady state, and the first instability occurs at wavelength $1/q \sim H^*$.

\subsection{Cell division, cell apoptosis, and tissue viscosity} \label{s:result-cell composition}
      	
     Sec.~\ref{s:method-model2} stated that the viscosity in a tissue depends on the local tissue composition.  A derivation of the viscosity from a general hydrodynamic model of stratified epithelium tissue is presented in the Appendix, where it is assumed that the stress relaxation process has negligible contribution from the turnover of junction proteins, and cell division/apoptosis produce force distribution that dominates the active stress inside a tissue.  This approach generalizes the earlier theory of Ranft et al.~\cite{Ranft_10} by explicitly including the contribution from different types of cells in a tissue \cite{Lander_09, Ovadia_13}.  Our analysis shows that the viscosity $\eta$ in a tissue can be expressed as
\begin{equation} \label{viscosity}
	\eta = \frac{\mu \sigma _0}{\rho \left[ (r_S d_S - r_D \tilde{d}_D )\Lambda_S + r_D \tilde{d}_D \right] },
\end{equation}
where $\Lambda_S = \rho _S/\rho $, $\rho_S$ is the number density of proliferative cells and $\rho$ is the total cell number density.  $\mu$ is the elastic shear modulus of the tissue, $r_S$ is the cell division rate and $r_D$ is the cell apoptosis rate, $d_S$, $\tilde{d}_{D}$ are characteristic strengths of the force dipoles during cell division and apoptosis, respectively.   $\sigma_0$ characterizes the response of cell polarity to the stress in a tissue, cells in a tissue with small $\sigma _0$ are more likely to be polarized by mechanical stress.  Equation~(\ref{viscosity})  suggests that viscosity in a tissue depends on the local proportion of proliferative cells.  This supports the argument we made in Sec.~\ref{s:method-model2}.  				
					
	Since in our model all the proliferative cells are localized in the basal ($z < h_S$) region, $\Lambda_S = 1$ for $z < h_S$ and $\Lambda_S = 0$ for $z > h_S$.  The viscosity in the tissue can be expressed as
\begin{eqnarray}
\label{eq:viscosity}
\eta (z) = \left\{
                     \begin{array}{ll}
                      \frac{\mu \sigma _0}{\rho r_S d_S} = \eta _S, & z < h_S \\
                       & \\
                      \frac{\mu \sigma _0}{\rho r_D \tilde{d}_D}=\eta_D, & z > h_S.
                      \end{array}
              \right.
\end{eqnarray}
					    From Eq.~(\ref{eq:viscosity}), it is clear that the relative viscosity $\eta_{rel} \equiv \eta_D / \eta_S$ and the relative proliferative rate $\tilde{r} \equiv r_D / r_S$ are not independent.  In fact one can express $\eta _{rel}$ as 
\begin{equation} \label{vis_r}
	\eta_{rel} = \frac{\tilde{\eta}_0}{\tilde{r}},
\end{equation}										
where $\tilde{\eta}_0 \equiv d_S  / \tilde{d}_D  > 0$ is a dimensionless parameter. 
					
\begin{figure}[t]
\begin{center}
	\epsfxsize=6.5in \epsfbox{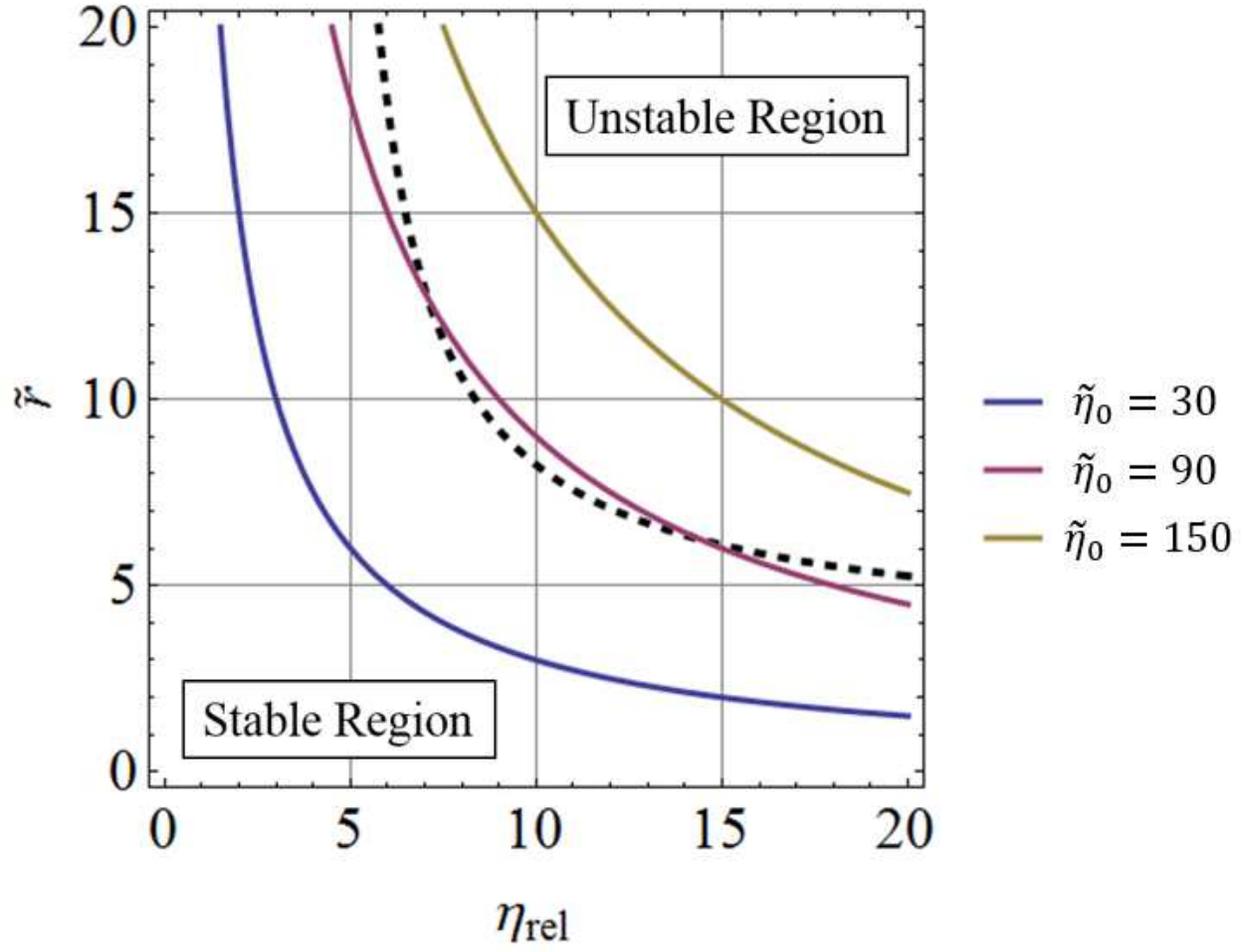}
	\caption{$\tilde{r}$ versus $\eta _{rel}$ for $\tilde{\eta}_0 =$ 30 (blue curve), 90 (red curve), and 150 (yellow curve).  The short-dashed curve is the stability boundary for a tissue with $\sigma q / (2 \eta_D r_D) = 0.2$ and $qH^\ast = \Xi$.  The stability boundary is determined from the zero of the relaxation rate for perturbations with the most dangerous wavelength.  }
\label{f:eta-r}
\end{center}				
\end{figure}

					 Equation~(\ref{eq:viscosity}) indicates that, for given $\tilde{\eta}_0$, for a tissue to have large $\eta _{rel}$, the magnitude of $\tilde{r}$ needs to be small.  On the other hand, Eq.~(\ref{ssc-model2}) suggests that as $\tilde{r}$ decreases, the proportion of proliferative region in the tissue also decreases.  Putting these two results together, Fig.~\ref{f:eta-r} shows $\eta _{rel}$ versus $\tilde{r}$ for a few choices of $\tilde{\eta} _0$.   For the contribution from $\omega _{phy}$ to drive the tissue to an instability, the magnitude of $\tilde{\eta}_0$ has to be large.  Thus a tissue with small $\tilde{d}_D$ is more likely to have a non-flat homeostasis state.
					
					Note that the viscosity shown in Eq.~(\ref{viscosity}) is derived for a tissue whose cell adhesion protein turnover has negligible contribution to its dynamics.  For a tissue in which the proliferative cells have small number of intercellular adhesion sites compared to terminal differentiated cells, $\eta _{rel}$ could be greater than the magnitude predicted by Eq.~(\ref{vis_r}).  Therefore when both turnover of cell adhesion proteins and active force dipoles due to cell division/cell apoptosis contribute significantly to tissue dynamics, one still should treat $\tilde{r}$ and $\eta _{rel}$ as independent parameters.

\section{CONCLUSION}\label{s:conclusion}		
		Although biological tissue exhibit high plasticity that allows remodeling, the homeostasis state is regulated such that its architecture is robust against intrinsic and external fluctuations~\cite{Lecuit_07, Lander_09}.  In this article we have studied how local tissue composition affects its viscous properties, and how viscous properties in turn affects the dynamics of a tissue close to a steady state of a stratified epithelium.   For a tissue with proliferative cells located close to basal membrane, and differentiated cells located close to apical surface, the lower region of the tissue has a different viscosity compared to the upper region.  Since cell division/apoptosis inevitably induces a flow in this tissue, the position-dependent viscosity in the tissue has a profound effect on its dynamics.  Our analysis shows that usually cell turnover is regulated to restore the steady state, but the flow induced by call apoptosis close to the apical surface induces a flow that hinder this relaxation process.  As a result, perturbations with wavelengths comparable to the thickness of the tissue has the slowest relaxation rate toward the steady state.  When the tissue has a thick proliferative region, and the viscosity of the proliferative region is significantly smaller than the other region, the steady state can even become unstable.   
		
		Our model is quite general, thus our prediction should hold in general for stratified continuous self-renewal tissues.  In the context of stratified epithelium, many undulating patterns, for example rete peg \cite{Hale_52, Babler_91, Lavker_83}, occur at basal lamina, and its mechanical origin has been explained by previous theoretical works \cite{Ciarletta_12, Basan_11, Ovadia_13, Risler_13}. On the other hand, our result provides a possible route to trigger epithelial apical surface undulations. Although an instability due to the induced flow described by our model is unlikely to happen in a normal stratified tissue, such mechanism could be important during embryogenesis and wound healing.   Furthermore, when two different tissues are in contact \cite{Basan_09}, the viscosity difference between these two tissues could also lead to interesting hydrodynamic instabilities.  These possibilities will be explored in our future works.  It is also worth mentioning that although in principle apical surface patterns can be a result of inhomogeneous distribution of growth factors in the tissue \cite{Ovadia_13}, our model reveal that pure mechanical force is sufficient to induce apical surface patterns.
	
		To test our theory by in vitro experiments, one could try to increase the death rate of differentiated cells to increase $\tilde{r}$, and decrease the tissue apical surface tension to make the contribution from 
$\omega _{phy}$ relatively more significant.  In the case of mammalian olfactory epithelium, a higher death rate can be induced by unilateral olfactory bulbectomy \cite{Holcomb_95}. The surface tension origins from the cell-cell adhesion \cite{Gonzalez-Rodriguez_12, Lecuit_07}, and it could be decreased by, for example the protease digestion procedure \cite{Weinberg_Cancer}.

		Several modifications can be made to make our model more general.  For example, the solid stroma in our model can be easily replaced by a soft stroma.  
Another important improvement that we will make in the future work is to develop a model in whch cell distribution is not pre-defined.  It is important to check how homeostasis state and pattern formation is achieved in a stratified epithelium with a more general hydrodynamic model.  The dynamics of cell lineage~\cite{Chou_10, Ovadia_13, Hannezo_14} will have to be taken into account in this future work.    

\section*{Acknowledgments}
The authors would like to thank support from the Ministry of Science and Technology, Taiwan (grant number 102-2112-M-008-008-MY3) and NCTS.   The authors are also grateful for stimulating discussions with J-F Joanny (ESPCI), and helpful discussions with C-M Chen (National Yang-Ming University, Taiwan).

\appendix
\section{$K_{phy}$ and $K_{mech}$}
\setcounter{equation}{0}			
\subsection{$K_{phy}$ and $K_{mech}$}
The analytical form of $K_{phy}$ and $K_{mech}$ in Section 2.2 are
\begin{equation}\label{dis_mech}
 K_{mech} \nonumber 
 = \frac{(\eta_{rel}+1)^2 (2qH^\ast-\sinh qH^\ast)+(\eta_{rel}-1)^2 \mu_1(qH^\ast,\tilde{h}_S)}{(\eta_{rel}+1)^2 (1+2(qH^\ast)^2+\cosh 2qH^\ast) + (\eta_{rel}-1)^2 \mu_2(qH^\ast,\tilde{h}_S)},
\end{equation}
\begin{equation}\label{dis_unsphy}
	K_{phy} = \frac{(\eta_{rel}+1)^2 [2(qH^\ast)^2]+(\eta_{rel}-1)^2 \mu_3(qH^\ast,\tilde{h}_S)}{(\eta_{rel}+1)^2 [1+2(qH^\ast)^2+\cosh 2qH^\ast] + (\eta_{rel}-1)^2 \mu_2(qH^\ast,\tilde{h}_S)}-1,
\end{equation}
where
\begin{equation}
\begin{split}
	\mu_1 = &  2qH^\ast \{1 -2\tilde{h}_S  [1+2(qH^\ast)^2](\tilde{h}_S-1)\tilde{h}_S \}+ 4qH^\ast(\tilde{h}_S-1) \cosh(2qH^\ast \tilde{h}_S) - \sinh[2qH^\ast(1-2\tilde{h}_S)]
	\\ & +2[1+2(qH^\ast)^2\tilde{h}_S^2]\sinh[2qH^\ast(1-\tilde{h}_S)],
\end{split}  
\end{equation}
\begin{equation}
\begin{split}
		\mu_2 = & 1 + 2(qH^\ast)^2 \{1+4(\tilde{h}_S-1)\tilde{h}_S[1+(qH^\ast)^2(\tilde{h}_S-1)\tilde{h}_S]\} 
		 	+ \cosh[2qH^\ast (1-2\tilde{h}_S)] 
		 	\\ & -2 [1+2(qH^\ast)^2\tilde{h}_S^2] \cosh[2qH^\ast(\tilde{h}_S-1)]  -2 [1+2(qH^\ast)^2(\tilde{h}_S-1)^2]\cosh(2qH^\ast \tilde{h}_S),
\end{split}
\end{equation}
\begin{equation}
	\mu_3 = 2(qH^\ast)^2 \{1+2\tilde{h}_S [-2+(1+2(qH^\ast)^2(\tilde{h}_S-1)^2)\tilde{h}_S]\}+2(\tilde{h}_S-1)^2\cosh(2qH^\ast \tilde{h}_S).
\end{equation}	
Here $\tilde{h}_S \equiv h_S/H^\ast$.  Although these expressions are complicated, they can be obtained from straightforward algebra.

\section{viscosity in a tissue}
In this Appendix we first construct a general model for continuous self-renewing tissues that includes both spatial and cell lineage dynamics. By assuming that cell division and apoptosis dominates the stress relaxation in a tissue, we show that a tissue behaves as a viscoelastic material, and the viscosity in a tissue depends on local tissue composition. The parameters in the theory can be estimated from existing experimental data of model systems.  The analysis in this Appendix follows closely the work of  Ranft et al.~\cite{Ranft_10}.  The main difference is that in our model the cell lineage, which has been identified as the fundamental unit of tissue and organ development \cite{Lander_09}, is also taken into account.
    	    	
    	There are proliferative cells and terminal differentiated (TD) cells in a tissue.   Typically, proliferative cells include stem cells and transit-amplifying (TA) cells differentiated from stem cells. 
 For simplicity we ignore this internal conversion of proliferative cells, only mark that proliferative cells can undergo cell division, and the daughter cells have certain chance to become terminal differentiated cells.    On the other hand, TD cells do not divide, they only undergo programmed cell death (apoptosis).  Similar to other models \cite{Lander_09, Chou_10, Ovadia_13}, we neglect apoptosis of proliferative cells.  
    	
    	Because of the coupling between the cortical network and the adhesion proteins, cells in a tissue are tightly connected~ \cite{Lecuit_07, Perez-Moreno_03}, and diffusion of cells inside the tissue can be  neglected \cite{Friedl_09}.  Denote $\rho _S$ as the number density of proliferative cells, $\rho_D$ as the number density of TD cells.  Taking into account cell division/apoptosis, and advection of cells by the flow inside the tissue, the continuity equations for $\rho _S$ and $\rho _D$ can be written as
\begin{equation}\label{cont_stem}
	\partial_t \rho_S + \partial_l (v_l \rho_S) = r_S (2 p_S - 1) \rho_S,
\end{equation}
\begin{equation}\label{cont_TD}
 	\partial_t \rho_D + \partial_l (v_l \rho_D) = 2 r_S (1 - p_S) \rho_S - r_D \rho_D,
\end{equation}
where Einstein summation convention is used to simplify the notation, $r_S$ and $r_D$ are the division rate of proliferative cells and apoptosis rate of TD cells, respectively.  
$p_S$ is the self-renewal probability of proliferative cells, and $v_i$ ($i=x, y, z$) is the $i$-th component of the flow field.  
    	
    	The total cell density at position ${\bf r}$ is $\rho ({\bf r},t) = \rho_S ({\bf r},t)+ \rho_D({\bf r}, t)$.  The fraction of the proliferative cells at ${\bf r}$ is   $\Lambda_S ({\bf r},t) = \rho_S ({\bf r}, t)/\rho({\bf r},t)$.  From Eq.~(\ref{cont_stem})(\ref{cont_TD}), the evolution equations for $\rho$ and $\Lambda _S$  are
\begin{equation}\label{continuity equation}
      	D_t  \rho = \rho [(r_S+r_D)\Lambda_S-r_D - \partial_l v_l],
\end{equation}
\begin{equation}\label{dynamic of cell lineage}
      	D_t \Lambda_S =[2 r_S p_S-r_S+r_D-(r_S+r_D)\Lambda_S]\Lambda_S,
\end{equation}
where $D_t \equiv \partial_t + v_l \partial_l$ is the material derivative.
    	
		We assume that, in the absence of active processes such as cell division and cell apoptosis, a tissue behaves as a linear elastic material, thus there is a linear relation between the change of elastic stress $\Delta \sigma^E_{ik}$ and the change of elastic strain $\Delta u_{ik}$, $\Delta \sigma^E_{ik} = (K - 2\mu/3) \Delta u_{ll}\delta_{ik}+2\mu \Delta u_{ik}$, where $K$ is the compressional modulus, and $\mu$ is the shear modulus~\cite{Landau_Elasticity}. Besides elastic stress, cell division and cell apoptosis also exert forces to the tissue~\cite{Kim_10}.  Because these processes exert no net force and torque on the tissue~\cite{Schwarz_02, Schwarz_02_2}, we model them as symmetric force dipoles~\cite{Ranft_10, Schwarz_02, Bischofs_04}. For a cell division/apoptosis event occurring at $\textbf{r} = \textbf{r}_0$, the change of stress is related to the active force dipole $d_{ik}$ through $\Delta \sigma^A_{ik} = -d_{ik} \delta(\textbf{r}-\textbf{r}_0)$, where $\delta(\textbf{r}-\textbf{r}_0)$ is the Dirac delta function, and superscript ``\textit{A}'' denotes stress created by active cell division/apoptosis events. Since inertia is negligible~\cite{Rauzi_11}, force balance condition gives
\begin{equation} \label{force_balance}
			\partial_i \sigma_{ik} = 0,
\end{equation}		    	
where $\sigma_{ik} \equiv \sigma_{ik}^E + \sigma_{ik}^A$ is the total stress in the tissue. 
    	
		Due to cell division and apoptosis, a unique reference state for the strain does not exist.  However, the difference of strain between subsequent states still can be defined \cite{Ranft_10}. Let the frame of reference flow and rotate with local flow and vortex, the evolution equation for the stress can be written as
\begin{equation} \label{Jaumann}
	D_t^J \sigma_{ik}= (K-\frac{2}{3}\mu)v_{ll} \delta_{ik}+2\mu v_{ik} + D_t^J \sigma^A_{ik},
\end{equation}		  	
where $D_t^J\sigma_{ik} \equiv D_t \sigma_{ik} + \Omega_{il} \sigma_{lk} + \Omega_{kl} \sigma_{il}$ is the Jaumann derivative. Here $v_{ik} \equiv (\partial_i v_k + \partial_k v_i)/2$ is the strain rate tensor, and $\Omega_{ik} \equiv (\partial_i v_k -\partial_k v_i)/2$ is the vorticity tensor.
    	
    	It is convenient to decompose the total stress tensor into isotropic and traceless parts, i.e.,  $\sigma_{ik} = -p \delta_{ik} + \tilde{\sigma}_{ik}$, where $p \equiv -\sigma_{ll}/3$ is the tissue pressure, and $\tilde{\sigma}_{ik}$ is the traceless (deviator) part of total stress tensor. Physically the isotropic part of the stress tensor is related to the change of local tissue volume, and $\tilde{\sigma}_{ik}$ is related to the distortion of local tissue shape~\cite{Landau_Elasticity}.  The same decomposition can also be applied to active stress tensor and strain rate tensor, i.e.~$\sigma_{ik}^A = -p^A \delta_{ik} + \tilde{\sigma}^A_{ik}$ and $v_{ik} = (1/3)v_{ll}\delta_{ik}+\tilde{v}_{ik}$.  Now the stress evolution Eq.~(\ref{Jaumann}) can be expressed as 
\begin{equation} \label{iso_dym}
    		D_t p = -Kv_{ll} + D_t p^A,
\end{equation} 
and
\begin{equation} \label{traless_dym}
    		D_t^J \tilde{\sigma}_{ik} = 2 \mu \tilde{v}_{ik} + D_t^J \tilde{\sigma}^A_{ik}.
\end{equation}
    	
\begin{figure}[t]
\begin{center}
	\epsfxsize=6.5in \epsfbox{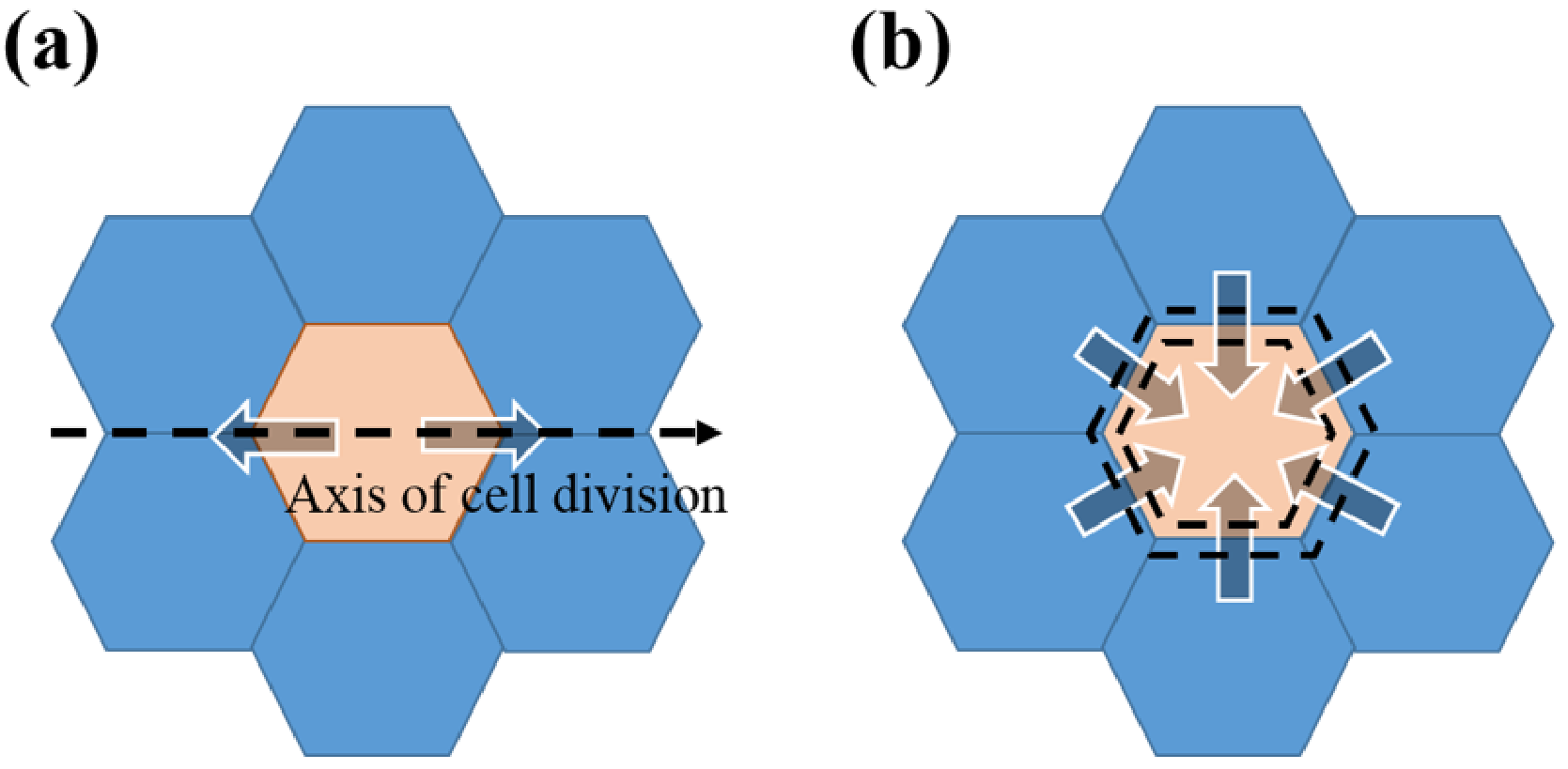}
	\caption{Cell division and apoptosis as a source of localized force inside tissue. (a) Cell division can produce equal but opposite forces along the mitotic axis (dashed arrow). (b) Cell apoptosis induce a purse-string (dashed line) which shrinks the cell surface and prevents gap formation. Both processes create a force dipole acting on the surrounding environment (blue cells).}
	\label{f:force_dipole}
\end{center}				
\end{figure}

		To complete our model, we still need the constitutive equations for $D_t p^A$ and $D_t^J \tilde{\sigma}_{ik}^A$.  Active stress is induced by cell division and cell apoptosis, we first consider the force dipole created by cell division.  Let unit vector $\hat{l}_i$ denote the direction of the cell division mitotic axis (Fig.~\ref{f:force_dipole}), follow~\cite{Tanimoto_12}, the force dipole produced by a cell division event can be approximated by 
\begin{equation} \label{single_division}
			d_{ik} ^{\text{div}} = 3d_S \hat{l}_i \hat{l}_k.
\end{equation}
Here $d_S$ characterizes the strength of the dipole. 
Typically $d_S > 0$ because cell division makes tissue grow.  The magnitude of $d_S$ can be deduced from experiments.  
Since different dividing cells may have different mitotic axis, the macroscopic properties is related to the average of $d_{ik}$ in a small region.  
\begin{equation} \label{division}
		\langle d_{ik}^{\text{div}} \rangle = 3d_S \langle \hat{l}_i \hat{l}_k \rangle = d_S (\delta_{ik} + \tilde{q}_{ik}),
\end{equation}
where $\tilde{q}_{ik} \equiv \langle 3\hat{l}_i \hat{l}_k -\delta_{ik}\rangle$ is the nematic order tensor in liquid crystal literatures.   
In general, tissue growth and cell division/apoptosis processes are intrinsically anisotropic \cite{Perez-Moreno_03, Rauzi_08, Bryant_08, Bittig_08, Ranft_10}, and the orientation of the cell motitic axis depends on local tissue distortion~\cite{Ranft_10, Thery_07, Marel_14}. To linear order, this relation can be expressed as
\begin{equation} \label{distortion}
			\tilde{q}_{ik} - \tilde{q}_{ik}^0 = \frac{\tilde{\sigma}_{ik}}{\sigma_0},
\end{equation}
where $\sigma_0 > 0$ is a constant, and $\tilde{q}^{0}_{ik}$ is the intrinsic anisotropic tensor, which should satisfy $\tilde{q}^{0}_{ll} = 0$. 
Denote the eigenvalues of $\tilde{q}_{ik}-\tilde{q}_{ik}^0$ as  $\tilde{q}^{(i)}$, it is convenient to write
\begin{equation}
\tilde{q}^{(i)} 
= 3 \langle \cos^2 \theta_i \rangle - 1,
\end{equation}
where $\theta_i$ is the direction angle of mitotic axis relative to the $i$-th principle axis of $\tilde{q}_{ik}-\tilde{q}_{ik}^0$. From Eq.~(\ref{distortion}) we have
\begin{equation} \label{order_pm}
	S_i = \frac{1}{6\sigma_0} \left( 3 \sigma^{(i)} - \sum_{k=1}^3 \sigma^{(k)} \right),
\end{equation}
where $S_i \equiv (3 \langle \cos^2 \theta_i \rangle - 1) / 2$ is the three-dimensional order parameter relative to $i$-th principle axis of $\tilde{q}_{ik}-\tilde{q}_{ik}^0$, and $\sigma^{(i)}$'s are the $i$-th eigenvalue of $\sigma_{ik}$. 
From this relation, one can estimate $\sigma_0$ from experimental measurements~\cite{Marel_14}.
		
		Cell apoptosis, a key mechanism of tissue morphogenesis~\cite{Toyama_08, Monier_15},  also exert force on the surrounding environment~(Fig.~\ref{f:force_dipole}). During apoptosis, a cell rapidly develops an actomyosin ring around its periphery and signal to neighboring cells to induce ``purse-string''-like contractility in the neighboring cells~\cite{Rosenblatt_01}. This ``purse-string'' contraction depends on myosin activity, which can be elevated by applying mechanical forces to the tissues \cite{Fernandez-Gonzalez_09}.  Therefore, in general the force dipole exerted by cell apoptosis is anisotropic and dependent on the local stress. 
\begin{equation} \label{apoptosis}
\langle d_{ik}^{\rm apo} \rangle = d_D\delta_{ik} + \tilde{d}_{D} \left( \tilde{q}^0_{ik} + \frac{\tilde{\sigma}_{ik}}{\sigma_{0}} \right) ,
\end{equation}
the traceless part of this force dipole has an intrinsic contribution proportional to $\tilde{q}^0_{ik}$ and a contribution induced by local stress. $d_D$ and $\tilde{d}_D$ are constants characterizing the strength of the isotropic and the traceless part of the dipole. Typically $d_D < 0$ since cell apoptosis often leads to the decrease of tissue volume.  On the other hand, there is no simple intuitive argument  to determine the the sign of $\tilde{d}_D$.
		
	Substitute Eq.~(\ref{division})(\ref{apoptosis}) into  Eq.~(\ref{iso_dym})(\ref{traless_dym}) leads to 
\begin{equation} \label{p_dym}
	D_t p = -K v_{ll} + \rho [(r_S d_S - r_D d_D) \Lambda_S + r_D d_D],
\end{equation}
\begin{equation} \label{Maxwell}
	(1+\tau D_t^J)\tilde{\sigma}_{ik} = 2\eta \tilde{v}_{ik} + \tilde{\sigma}^I_{ik},
\end{equation}
where $\tilde{\sigma}^I_{ik} \equiv \tau \rho [(r_S d_S - r_D \tilde{d}_D)\Lambda_S+r_D \tilde{d}_D]\tilde{q}^0_{ik}$ is the intrinsic active stress due to cell intrinsic polarity, $\eta \equiv \tau \mu$, and the time scale $\tau$ that characterizes the active stress induced by cell proliferation/apoptosis is given by 
\begin{equation} \label{tau}
	\tau^{-1} \equiv \frac{\rho}{\sigma_0} [(r_S d_S-r_D \tilde{d}_D)\Lambda_S + r_D \tilde{d}_D].
\end{equation}
		
From Eq.~(\ref{Maxwell}) and Eq.~(\ref{tau}), we make the following remarks.
\begin{itemize}
\item For a tissue with no intrinsic cell polarity, $\tilde{\sigma}^I_{ik} = 0$.  The stress evolution  Eq.~(\ref{Maxwell}) reduces to the Maxwell model of a visco-elastic material with stress relaxation time $\tau$ and viscosity $\eta$.  At time scale greater than $\tau$, the tissue behaves like a fluid \cite{Mase_ContiMech},  
\begin{equation}\label{fluid equation}
	   \tilde{\sigma}_{ik} = 2\eta \tilde{v}_{ik}.
\end{equation}
It is important to note that the viscosity of the tissue depends on $\rho$ and $\Lambda_S$.  This 
supports the intuitive argument that we presented in the main text that the tissue viscosity depends on 
the local density of proliferative cells and TD cells.
				
\item From Eq.~(\ref{tau}) a sufficient condition for the tissue viscosity to be positive is $\tilde{d}_D>0$. This is consistent with what has been observed in the experiments \cite{Foty_94,Forgacs_98,Bonnet_12}.
			
\item In general $\tilde{\sigma}^I_{ik}$ is not zero. In this case, in the long time limit the stress of the tissue satisfies
\begin{equation}
		\tilde{\sigma}_{ik} = 2\eta \tilde{v}_{ik} + \tilde{\sigma}^I_{ik}.
\end{equation}
The additional term on the right-hand side of the above equation suggests that a tissue in a flow-free state is in general under stress.  This is the active stress from cells with intrinsic polarity.  It is important for tissue morphogenesis, for example, in the imaginal disk of \textit{Drosophila} \cite{Ranft_10,Bittig_08}.
		\end{itemize}
		
	At time scale large compare to  $\tau $, the total stress tensor in a tissue reduces to	
\begin{equation} \label{long-term}
	\sigma_{ik} = -p \delta_{ik} + 2 \eta \tilde{v}_{ik} + \tilde{\sigma}^I_{ik}.
\end{equation}
	Hence a tissue in this limit behaves as a viscous fluid coupled to a scalar field $\Lambda_S$ and an order parameter which measures the intrinsic anisotropy of cell orientation.  Moreover, Eq.~(\ref{long-term}), Eq.~(\ref{continuity equation}-\ref{force_balance}) and Eq.~(\ref{p_dym}) give a close set of equations describing tissue and cell lineage dynamics.  Note that the constitutive relation we used in the main text describes a tissue with negligible intrinsic active stress.
		
	 By measuring traction stress exerted by the dividing \textit{Dictostelium} cell on the flexible poly-acrylamide gel, Tanimoto and Sano~\cite{Tanimoto_12} found that the typical strength of force dipole during cell division is about $1.2\times10^{-13}~\text{N}\cdot\text{m}$ and the orientation of the force dipole coincides with cell division axis \cite{Tanimoto_12}.  This suggests that $d_S $ in Eq.~(\ref{single_division}) can be measured.  Although  \textit{Dictostelium} does not form tissues, one can use this experimental measurement to get a rough idea of the order of magnitude of $d_S$ as $d_S \sim 4\times10^{-14}~\text{N}\cdot\text{m}$.
	
	 Experiments similar to~\cite{Marel_14} can be applied to measure $\sigma _0$.  In this experiment, Marel et al. measured the average cell division orientation as a function of local strain rate in monolayer Madin-Darby canine kidney (MDCK) cells and concluded that 
\begin{equation} \label{2dexp_Marel}
		2 \langle \cos^2 \phi\rangle - 1 \cong \frac{5}{8} (\lambda_1 - \lambda_2),
\end{equation}
where $\lambda_1$ and $\lambda_2$ ($\lambda _1 > \lambda _2$) are the eigenvalues of strain rate tensor $v_{ik}$, and $\phi$ is the angle between cell division axis and the principle axis of the eigenvalue $\lambda_1$. By assuming the monolayer tissue behaves as a Newtonian fluid, we can use the two-dimensional form of Eq.~(\ref{order_pm}) , 
\begin{equation} \label{2d-order_pm}
	2 \langle \cos^2 \phi\rangle - 1 \cong \frac{\bar{\eta}}{\bar{\sigma}_0} (\lambda_1 - \lambda_2),
\end{equation}
where $\bar{\eta}$ is the 2-dimensional viscosity. Using typical tissue viscosity $\eta \cong 10^5~\text{Pa}\cdot\text{s}$ \cite{Forgacs_98} and MDCK monolayer height $h \cong 2~\mu\text{m}$ \cite{Hoh_94}, a comparison between Eq.~(\ref{2dexp_Marel}) and Eq.~(\ref{2d-order_pm}) gives $\bar{\sigma}_0 \cong 5~\text{Pa}\cdot\text{m}$. Thus $\sigma_0 = \bar{\sigma}_0 / h \cong 2.5 \times 10^6~\text{Pa}$.			

              For amnioserosa tissue, the magnitude of $\tilde{d}_D$ can be inferred from the apoptosis rate measured from dorsal closure process of \textit{Drosopila}. The  measurement made by Toyama et al. gives $r_D \cong 5.5\times10^{-5}~\text{s}^{-1}$~\cite{Toyama_08, footnote}. 
Since during dorsal closure, amnioserosa cell only undergo cell apoptosis,  Eq.~(\ref{tau}) is reduced to $\tau ^{-1} = \rho r_D \tilde{d}_D/\sigma _0$.  Use tissue height $h \cong 5 \mu\text{m}$ \cite{Saias_15}, cell density $\rho \cong 2.2\times 10^{-3}~\mu\text{m}^{-3}$ \cite{Toyama_08}, and the relaxation time $\tau \cong 100~\text{s}$ \cite{Solon_09}.  If we use the magnitude of $\sigma _0$ measured from MDCK cells, we find $\tilde{d}_D \cong 2.1\times10^{-7}~\text{N}\cdot\text{m}$ for amnioserosa tissue.   In principle, combining the measurement by~\cite{Toyama_08} with an experiment like ~\cite{Marel_14} for amnioserosa tissue give us a better estimate for its $\tilde{d}_D$.

In summary, in this Appendix we have presented a theoretical model to relate the viscosity of a tissue to a few measurable parameters.  The key parameters $d_S$, $\tilde{d}_D$, and $\sigma _0$ of a tissue can be measured from different experiments.  Therefore it is possible to infer the hydrodynamic properties of a tissue from existing experimental techniques.

\newpage
			    	
 \begin{figure}[t]
	\begin{center}
	\epsfxsize= 6.5 in \epsfbox{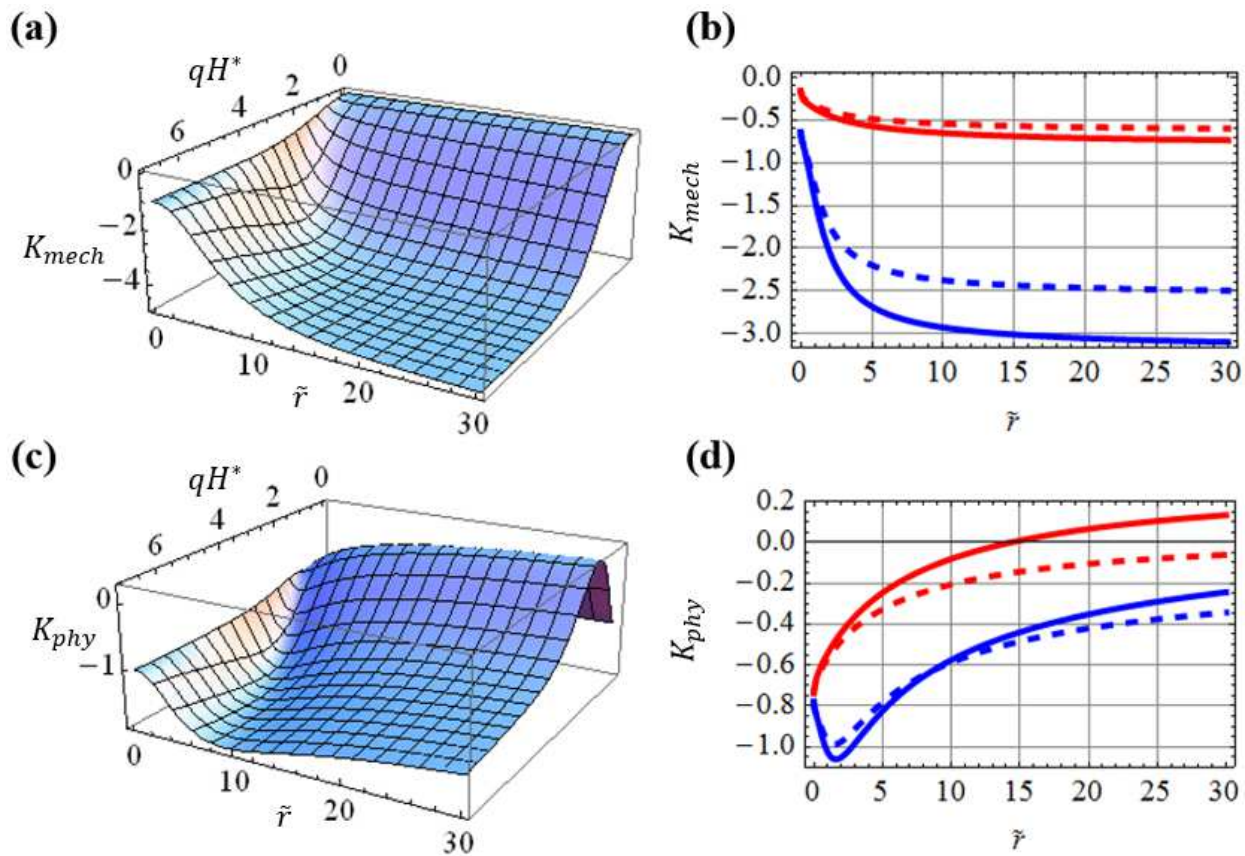}	
		\caption{ (a) $K_{mech}$ as a function of $\tilde{r}$ and $qH^\ast$ for $\eta _{rel} = 5$. (b) $K_{mech}$ versus $\tilde{r}$ for  $qH^\ast = 2$, $\eta_{rel} = 5$ (solid blue curve), $qH^\ast = 0.6$, $\eta_{rel} = 5$ (solid red curve),  $qH^\ast = 2$,  $\eta_{rel} = 4$ (dashed blue curve), and  $qH^\ast = 0.6$,  $\eta_{rel} = 4$ (dashed red curve). (c)  $K_{phy}$ as a function of $\tilde{r}$ and $qH^\ast$ for $\eta_{rel} = 5$. (d)  $K_{phy}/r_D$ versus $\tilde{r}$  for   $qH^\ast = 2$, $\eta_{rel} = 5$ (solid blue curve), $qH^\ast = 0.6$, $\eta_{rel} = 5$ (solid red curve),  $qH^\ast = 2$,  $\eta_{rel} = 4$ (dashed blue curve), and  $qH^\ast = 0.6$,  $\eta_{rel} = 4$ (dashed red curve).  }
		\label{f:new_mech_phy}
	\end{center}				
\end{figure}	
	    
\begin{thebibliography}{10}
\bibitem{footnote}
Note that the apoptosis rate measured in \cite{Toyama_08} corresponds to $N_{\text{total}}$ times $r_D$ in our model.   $N_{\text{total}}$ is the number of amnioserosa cells just before dorsal closure. Typically $N_{\text{total}} \approx 100$.
\end{thebibliography}
\end{document}